# Experimentally demonstration of the repulsive Casimir force in the gold-cyclohexane-PTFE system

Qian Hu†, Jianhua Sun†, Qian Zhao*, and Yonggang Meng

[1]State Key Laboratory of Tribology, Department of Mechanical Engineering,
Tsinghua University, Beijing, 100084, China

E-mail: zhaoqian@mail.tsinghua.edu.cn

† contributed equally to this work.

**ABSTRACT**

The experimentally demonstration of Casimir force transition from attraction to repulsion is still challenging. Herein, the Casimir forces for a sphere above a plate immersed in different liquids were precisely measured using Atomic force microscope, and the long-range repulsive Casimir force in the gold-cyclohexane-PTFE system is observed for the first time. The experimental data are consistent with the calculation by Lifshitz theory, which offers the direct evidence for the system of $\varepsilon_1<\varepsilon_3<\varepsilon_2$. It further verifies the reasonability of van Zwol *et al.* dielectric model to describe the intervening fluids. This study is promising for potential applications on quantum levitation and frictionless devices in MEMS and NEMS by Casimir repulsion.

In 1948, Casimir predicted that there would be an attractive force between two electrically neutral metallic plates in vacuum, which was named as the Casimir effect [1]. According to quantum electrodynamics, the force arises from electromagnetic fluctuations created by quantum [2]. Following this work, Lifshitz generalized the Casimir force between arbitrary dielectric plates [3]. In Lifshitz theory, the retardation effects, owning to the finite speed of light, lead to a subtle distinction between the Casimir and the familiar van der Waals forces; the former works at larger distances [4].

Lifshitz theory indicates that the dielectric response functions of the materials directly affect the Casimir force between the surfaces. Therefore, it's essential to acquire the dielectric function of the fluid over a wide frequency range for the theoretical force. The acquisition of dielectric functions often requires a very amount of data, its measurement methods are difficult and the accuracy is not high. So far, different theoretical models for fluids are created to describe their dielectric functions, however, there has been no inconclusive ideas on which dielectric model is the most suitable for a certain fluid. It is urgent for us to use another method to demonstrate the validity of the dielectric functions of the fluids, like using experiment results to indirectly verify.

Generally, the Casimir force is attractive, and the existence of such an attraction may give rise to the irreversible adhesion of neighboring elements in micro- and nanoelectromechanical systems (MEMS and NEMS) [5-8]. It can be seen from Lifshitz theroy that Casimir repulsion will occur when two interacting objects are separated by a certain medium and the term $\varepsilon_1 < \varepsilon_3 < \varepsilon_2$ could be satisfied, where $\varepsilon_1$, $\varepsilon_2$ and $\varepsilon_3$ are the dielectric response functions of the two interacting bodies and the medium, respectively [9]. According to the existing studies on the dielectric functions of various materials, it has been elusive to obtain Casimir repulsion in the atmosphere or other gases. This is because it's difficult to find any gases with a higher dielectric constant than a certain solid material. Nevertheless, by introducing fluids with a high refractive index, Lifshitz's term could be satisfied for a few interacting pairs [10]. In this way, the long-range Casimir repulsive force was first measured between gold and silica surfaces in bromobenzene [11]. Very recently, a stable Casimir equilibrium has been demonstrated between gold nanoplate and Teflon-coated gold surface in ethanol caused by the co-effect of repulsion at short separations and attraction at long separations [12]. On the other hand, according to Boyer's theory, Casimir repulsion can also occur between a primarily electric object and a primarily magnetic object [13]. As a result, metamaterials are the key to achieving Casimir repulsion, such as chiral metamaterials [14-16], topological metamaterial [17,18] and electromagnetic metamaterials [19-22]. Of course, there are many studies that measure the van der Waals repulsion, similar to the Casimir repulsion. Adam *et al.* [23] measured van der Waals repulsive force between a gold sphere and a flat PTFE surface in cyclohexane using colloid probe atomic force microscopy, however, most of the data they measured was limited to separation distances below 20 nm, and we still knew nothing about the long-range Casimir force. In this letter, we have tried to exclude the effect of electrostatic force, hydrodynamic force and roughness. Then we present precise measurements of the Casimir force between a gold sphere and a flat plate separated by a liquid medium. Finally, we compare the experiment results with the theoretical analysis from different dielectric models of the liquid and discuss which one is the most reasonable.

According to Lifshitz theory, the Casimir force between a sphere and a plate immersed in a fluid separated by a distance $d$ can be expressed as [3]:

$$F(d) = \frac{\hbar R}{2\pi c^2} \int_1^\infty p\, dp \int_0^\infty \varepsilon_3 \xi^2 d\xi \left[ \ln\left(1 - \frac{s_1\varepsilon_3 - s_3\varepsilon_1}{s_1\varepsilon_3 + s_3\varepsilon_1} \frac{s_2\varepsilon_3 - s_3\varepsilon_2}{s_2\varepsilon_3 + s_3\varepsilon_2} e^{-\frac{2\xi pd\sqrt{\varepsilon_3}}{c}}\right) + \ln\left(1 - \frac{s_1 - s_3}{s_1 + s_3} \frac{s_2 - s_3}{s_2 + s_3}\right) e^{-\frac{2\xi pd\sqrt{\varepsilon_3}}{c}} \right] \quad (1)$$

where $s_k = \sqrt{p^2 - 1 + \frac{\varepsilon_k}{\varepsilon_3}}$, $k = 1, 2, 3$, $\hbar$ is the reduced Planck constant, $c$ is the speed of light, and $\varepsilon_1$, $\varepsilon_2$ and $\varepsilon_3$ are the dielectric functions of the sphere, the plate and the fluid, respectively, evaluated at imaginary frequencies according to the Kramers-Kronig relation [3]:

$$\varepsilon_j(i\xi) = 1 + \frac{2}{\pi}\int_0^\infty \frac{\omega \text{Im}[\varepsilon_j(\omega)]}{\xi^2+\omega^2} d\omega \quad (2)$$

According to Equation (1), when Lifshitz theory is used to calculate Casimir force, dielectric model determines whether the theoretical prediction is accurate or not. However, many theoretical models for fluids are created to describe their dielectric functions, leading to some fluids with many dielectric models. For ethanol, there are three common dielectric models: the two-oscillator model of Milling *et al*, the three-oscillator model of van Oss *et al.* and the seven-oscillator model of van Zwol *et al*.

The two-oscillator model proposed by Milling *et al.* [24] predicts a repulsive force between gold and silica surfaces in ethanol, while an attractive force was experimentally observed in this system [25]. The two-oscillator model expressed as [24]:

$$\varepsilon(i\xi) = 1 + \frac{c_{IR}}{1+(\frac{\xi}{\omega_{IR}})^2} + \frac{c_{UV}}{1+(\frac{\xi}{\omega_{UV}})^2} \quad (3)$$

where $\omega_{IR} = 6.60 \times 10^{14}$ rad/s and $\omega_{UV} = 1.14 \times 10^{16}$ rad/s are the characteristic absorption frequencies in the infrared and ultraviolet range, respectively, and $c_{IR} = 23.84$ and $c_{UV} = 0.852$ are the corresponding absorption strengths.

Compared with the model of Milling *et al.*, the dielectric data in the microwave range is added to the model of van Oss *et al.*, thus forming a three-oscillator model [26]:

$$\varepsilon(i\xi) = 1 + \frac{\varepsilon_0 - \varepsilon_{IR}}{1+\frac{\xi}{\omega_{MW}}} + \frac{\varepsilon_{IR} - n_0^2}{1+(\frac{\xi}{\omega_{IR}})^2} + \frac{n_0^2 - 1}{1+(\frac{\xi}{\omega_{UV}})^2} \quad (4)$$

where $n_0$ is the refractive index in the visible range with $n_0^2 = 1.831$, $\varepsilon_0 = 25.07$ is static dielectric constant, $\varepsilon_{IR} = 4.2$ is the dielectric constant in the infrared range, and $\omega_{MW} = 6.97 \times 10^9$ rad/s, $\omega_{IR} = 2.588 \times 10^{14}$ rad/s, $\omega_{UV} = 1.927 \times 10^{16}$ rad/s are the characteristic microwave, infrared, and ultraviolet absorption frequencies, respectively. Different from the two-oscillator model, it predicts an attractive force, though not matching the magnitude of the measurement results [26].

Van Zwol *et al.* constructed the dielectric function of various materials from abundant literature data measured over a wide frequency interval, also based on the

oscillator model [27]:

$$\varepsilon(i\xi) = 1 + \sum_i \frac{C_i}{1+(\xi/\omega_i)^2} \quad (5)$$

where $C_i$ is the oscillator absorption strength at a given characteristic frequency $\omega_i$. For ethanol, seven oscillators are used, the detailed dielectric data is referred from [27]. At present, the model is used in many theoretical studies about Casimir force, but its reasonability needs further verification.

Similar to ethanol, cyclohexane also has two controversial dielectric models: the model of Milling *et al.* and the model of van Zwol *et al.* Here the model of Milling *et al.* is one-oscillator model [28]:

$$\varepsilon(i\xi) = 1 + \frac{c_{UV}}{1+(\frac{\xi}{\omega_{UV}})^2} \quad (6)$$

where $\omega_{UV} = 1.142 \times 10^{16}$ rad/s are the characteristic absorption frequency in the ultraviolet range, and $c_{UV} = 1.023$ is the corresponding absorption strength. The model of van Zwol *et al.* for cyclohexane has the form as the equation (5), the number of the oscillator is $i = 5$, referred from [27].

For solid materials, the existing dielectric model is more accurate and can be considered as not controversial. The dielectric function for gold is described by the Drude model:

$$\varepsilon(\omega) = 1 - \frac{\omega_p^2}{\omega^2 + i\gamma\omega} \quad (7)$$

where $\omega_p = 9.0$ eV and $\gamma = 0.035$ eV, referred from [29]. And the model of van Zwol *et al.* for silica and PTFE has the form as the equation (5), the numbers of the oscillator are $i = 8$, referred from [27].

We use a commercial atomic force microscope (AFM, Asylum Research MFP-3D) to perform the measurements of Casimir force between the gold microsphere and plate with the intervening liquids. A schematic diagram of the experimental setup is shown in Fig. 1(a). Considering that the Casimir force is very small, a sphere with a large volume is needed in order to amplify the force. As a result, a bigger spring constant of the cantilever is needed to support the gold sphere. However, the bigger spring constant will decrease the sensibility of measurement. Therefore, we chose a barium titanate sphere with smooth surface, low density and easy to sputter gold as the substrate, a sputter coater system (Leica EM ACE600) was used to sputter a gold film with a thickness of more than 100 nm, and the diameter of the sphere obtained by AFM was 74 μm. Then the barium titanate sphere is attached to the tip of a triangular cantilever (NP-O10) with epoxy adhesive. A scanning electron microscope (SEM) image of the cantilever with the coated sphere attached is shown in Fig. 1(a).

The gold surfaces are prepared by coating $15 \times 15$ mm$^2$ silicon wafers with 100 nm thickness of gold. The silica surfaces are directly from the commercial supplies ($10 \times 10$ mm$^2$ silica wafers). The PTFE surfaces are prepared by polishing the samples by Chemical Mechanical Polishing (CMP, UNIPOL-1000S). The intervening liquids, including ethanol and cyclohexane, are from the commercial supplies with the purity of $\geq 99.8\%$ (HPLC).

To precisely measure Casimir force, several related forces should be excluded, such as electrostatic force and hydrodynamic force etc.

First is how to exclude the effect of electrostatic force. It is not negligible in the measurement of the Casimir force in air, however, when in fluids, due to the induced dielectric polarization, the electrostatic force is reduced significantly [30]. A surface potential scan is performed on the gold plates by AFM before the measurements and the potential of the surfaces is below 60 mV, as shown in FIG.1(b). In this case, the electrostatic force is less than 10pN in fluid, which can be negligible compared with the Casimir force in our measurements.

Then, considering how to exclude the hydrodynamic force under the fluids. The hydrodynamic force is related to not only the properties of the material, but also the velocity *v* at which the sphere approaches the plate [31]:

$$F_{hydro}(d,v) = -\frac{6\pi\eta v}{d}R^2 \tag{8}$$

where $R$ is the radius of the sphere, $\eta$ is the viscosity of the fluid, $d \ll R$ is the distance between the sphere and the plate. The minus sign means the direction of the hydrodynamic force is opposite to that of the moving velocity. According to equation (8), the hydrodynamic force can be negligible at an extremely low velocity. To prove it, we obtain the measurement data at several constant piezo velocities using another sphere for the configuration of Au-ethanol-SiO$_2$. Fig. 1(c) indicates that the hydrodynamic force is dominant and increases with the decrease of the piezo displacement at velocities larger than 1 μm/s. This is qualitatively accordant with the equation (8). When the cantilever approaches at a velocity $v = 0.03$ μm/s, an attractive force (the Casimir force) occurs, almost free from the hydrodynamic force (See the Supplemental Material for more information).

In addition to electrostatic force and hydrodynamic force, roughness can also affect the precise measurement of Casimir force. We have found through experiments that hydrodynamic can only be measured in areas with very small roughness. Therefore, we chose these areas to measure Casimir force to indirectly eliminated the effect of roughness on the experiments. Fig. 1(d) shows that the topography images of gold plate scanned by AFM with a roughness of $R_a < 0.5$ nm. The roughness of SiO$_2$ and PTFE plates are below $0.5\ nm$ and $3.5\ nm$ respectively (see FIG S2 of the Supplemental Material). For the Casimir force to be studied at separations $d > 20$ nm, the corrections due to such a roughness can be omitted [32].

The experiment results for these configurations are shown in Fig. 2(a). When the sphere is far away from the plate, it is free from any forces, so the force curve is a flat segment. After the sphere enters into the working range of the Casimir force, the deflection of the cantilever increases gradually. For the configurations of Au-ethanol-Au and Au-ethanol-SiO$_2$, a negative deflection of the cantilever arises, implying an attractive force, then the sphere is pulled to contact the plate instantaneously. By contrast, the deflection data for Au-cyclohexane-PTFE is positive, implying a repulsive force, then the cantilever bends to the direction away from the plate but the sphere still will touch it driven by the piezoelectric column. Eventually, the force curve turns into a straight line with a fixed slope. With the curve parts at distances less than 20 nm removed to avoid the pull-in effect of the attractive force, Fig. 2(b) indicates that the force for Au-ethanol-Au is significantly larger than that for the other two configurations: at $d = 25$ nm, the measured Casimir force for Au-ethanol-Au is close to 1000 pN, while for Au-ethanol-SiO$_2$ and Au-cyclohexane-PTFE the absolute force is less than 100 pN.

Then we obtain the theoretical Casimir forces between the gold sphere and the silica plate immersed in ethanol calculated from the three different dielectric functions of ethanol, as shown in Fig. 3(a). The scatter plots represent four sets of experimental results in continuous runs. Apparently, the two-oscillator model leads to a repulsive force, while the three-oscillator model and the seven-oscillator model lead to attractive forces. Thus, the two-oscillator model for ethanol is unreasonable because it predicts an absolute converse sign on the Casimir force. For the other two models, the seven-oscillator model for the dielectric of ethanol predicts a stronger force than the three-oscillator model. It reveals that the measurement results are in reasonable agreement with Lifshitz theory calculated from the seven-oscillator model.

Figure.3(b) shows the case for Au-ethanol-Au. The three-oscillator model and the seven-oscillator model of ethanol predict almost an identical magnitude of forces, causing the two theoretical curves to overlap each other. By comparison, the two-oscillator model predicts a slightly weaker force. The measurement results have coincided well with the three-oscillator model and the seven-oscillator model.

Figure 4 shows experimental results for the configuration of Au-cyclohexane-PTFE with these two different calculations. The theoretical forces calculated from the two models for cyclohexane have a slight discrepancy at smaller distances. The model of Milling *et al.* predicts a slightly weaker repulsive force. By contrast, the measurement results coincide more with theoretical ones calculated from the model of van Zwol *et al.*

According to the analysis above, the dielectric model of van Zwol *et al.* for the fluids gives a more consistent theoretical Casimir force with the experimental results for all three configurations than the models of van Oss *et al.* and Milling *et al.* The simple oscillator models which generally have an oscillator number of less than or equal to 3 were constructed from limited measured dielectric data. Therefore, it may not predict an enough accurate dielectric function for the fluids in a wide frequency range. Further, the dielectric model of van Zwol *et al*, including more oscillators, carefully used more dielectric data from a lot of literatures. This explained why the model of van Zwol *et al.* predicts a Casimir force closer to the measurement results.

To further verify the reasonability of van Zwol et al. dielectric model, the dielectric functions of SiO2, PTFE, ethanol and cyclohexane based on van Zwol *et al.* dielectric model are shown in Fig.5. Lifshitz' term $\varepsilon_1 < \varepsilon_3 < \varepsilon_2$ indicates that when the dielectric function of the intervening medium is between those of the two interacting bodies, Casimir repulsion will occur. For Au-ethanol-Au, the two interacting surfaces consists of the same material, which can't be suitable for Lifshitz' term. In this case, the Casimir force is attractive with no doubt. And the force is rather large because the both surfaces are gold that is a good conductor and close to a perfect metal. For Au-ethanol-SiO$_2$, Lifshitz' term is satisfied only in a lower frequency range (approximately $< 2.5 \times 10^{12}$ rad/s), however, the dielectric functions in a higher frequency range have a crucial effect on the Casimir force at submicron separations we studied in this paper. Therefore, in this case, the force is attractive but extraordinarily weak due to that the dielectric function of ethanol is close to the dielectric function of silica in the frequency range. For Au-cyclohexane-PTFE, the dielectric function of PTFE is the lowest in all materials, creating a favorable condition to match Lifshitz' term in the frequency range. Only in this configuration, repulsive Casimir force occurs at all separations.

In conclusion, excluding the effects of electrostatic force, hydrodynamic force and roughness, we have performed precision measurements of the Casimir force for a sphere above a plate immersed in liquids by AFM. The long-range repulsive Casimir force in the gold-cyclohexane-PTFE system is observed for the first time, it is beneficial to promote the development of frictionless devices and reduce the loss of MEMS devices. Then we use an indirect method to compare the experimental data with Lifshitz theory calculated from different dielectric models for the intervening fluids, which is to verify the reliability of the dielectric model. We find the model of van Zwol *et al.* is the most consistent. The work eliminates the current controversy in the dielectric model and provides guidance for future theoretical research.

**Acknowledgements**

This work is supported by the National Natural Science Foundation of China (Grant Nos. 51575297, 51872154, 51635009), the Science and Technology Plan of Shenzhen City (JCYJ20170817162252290), and the Chinese State Key Laboratory of Tribology. The authors gratefully acknowledge discussions with Prof. J. Zhou, Dr. L. Kang, Dr. N.-H. Shen, and Dr. P. Zhang.

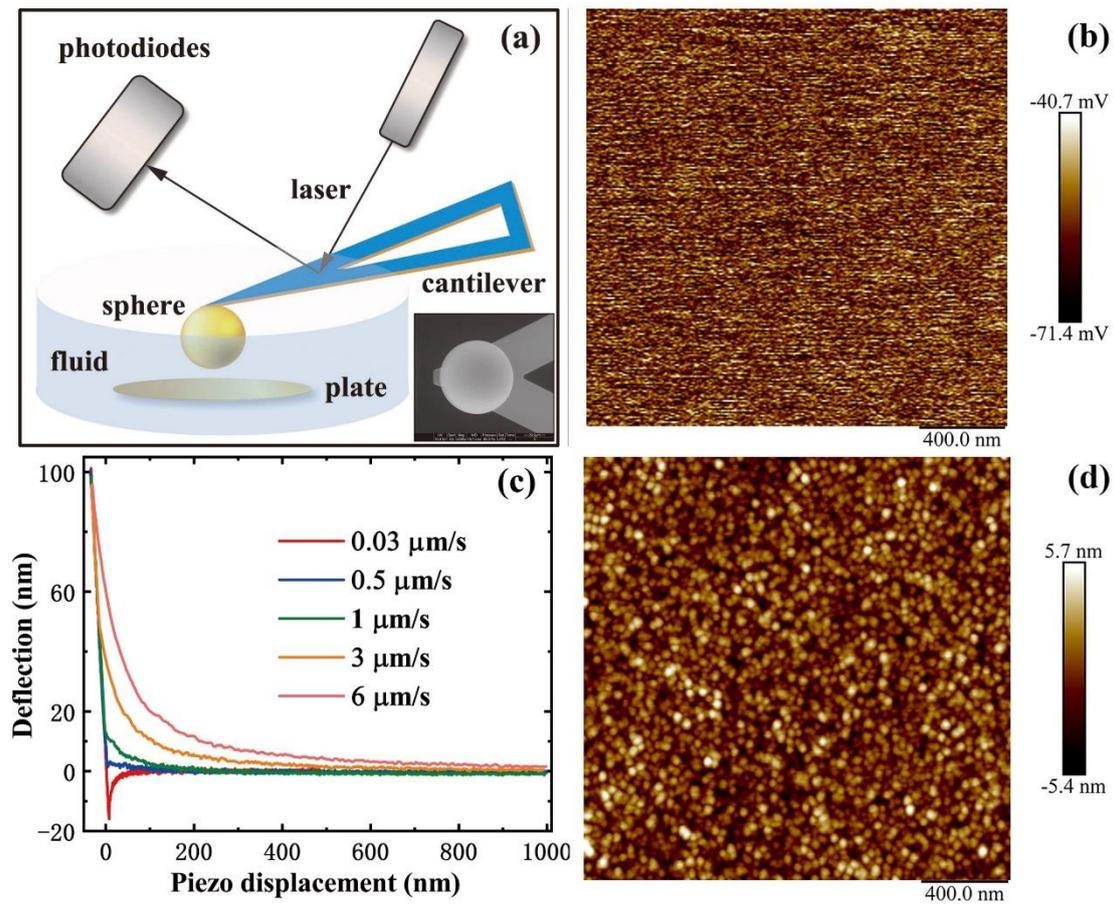

FIG. 1 (a) (Color online) Schematic of the experimental setup. Inset: SEM image of the gold coated sphere attached on an AFM cantilever. (b) The surface potential of the gold plate is below 60mV. (c) Raw deflection data versus piezo displacement at various approach velocities. (d) The surface topography images of gold plate is below 1.5 nm.

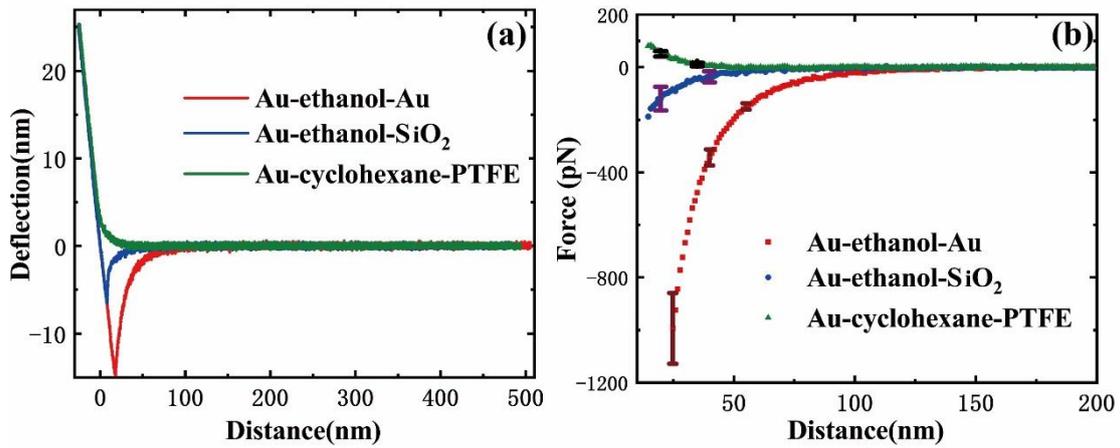

FIG. 2 (a) Raw deflection data versus piezo displacement for three configurations of Au-ethanol-Au, Au-ethanol-SiO$_2$ and Au-cyclohexane-PTFE. (b) Experimental results averaged by ten sets of data for these configurations.

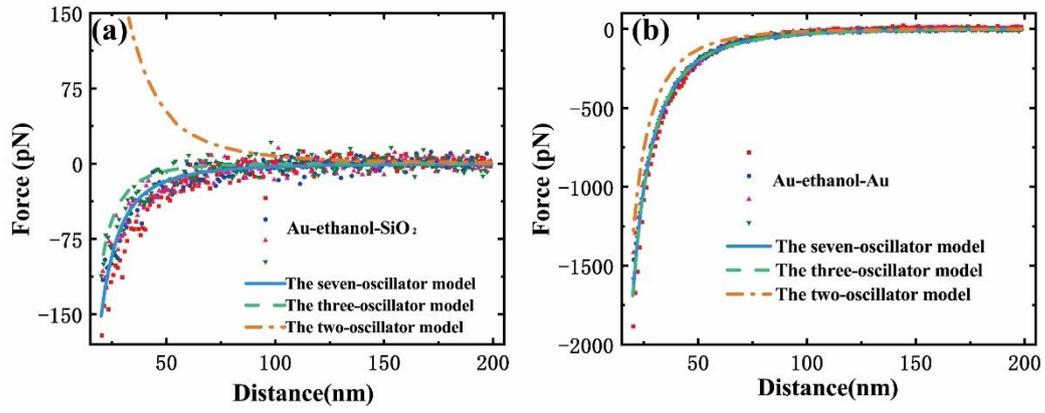

FIG. 3 (Color online) Four sets of experimental results for the configuration of (a) Au-ethanol- SiO2 and (b) Au-ethanol-Au with different calculations (solid, dashed and dashed dotted line) based on Lifshitz theory from three different dielectric models of ethanol.

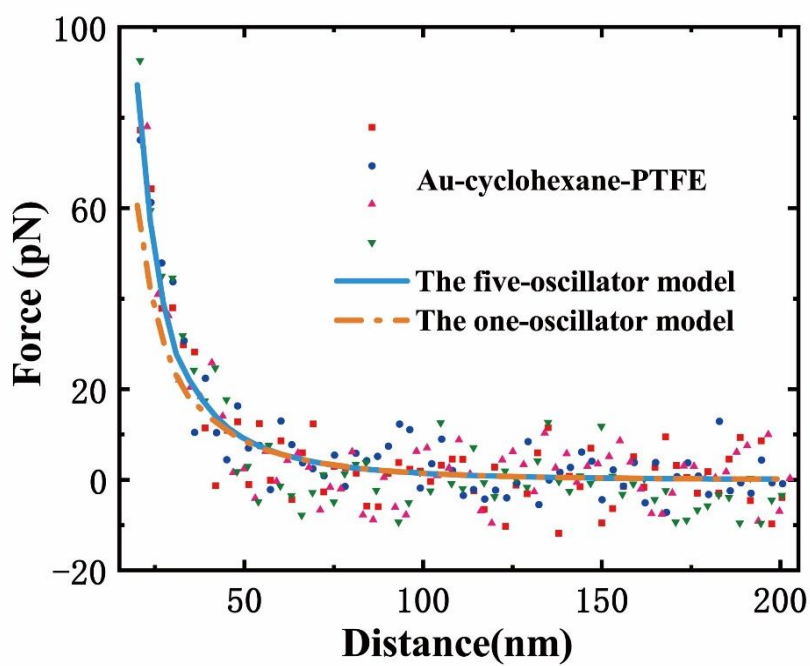

FIG. 4 (Color online) Four sets of experimental results for the configuration of Au-cyclohexane-PTFE with different calculations (solid line and dashed dotted line) based on Lifshitz theory from two different dielectric models of cyclohexane.

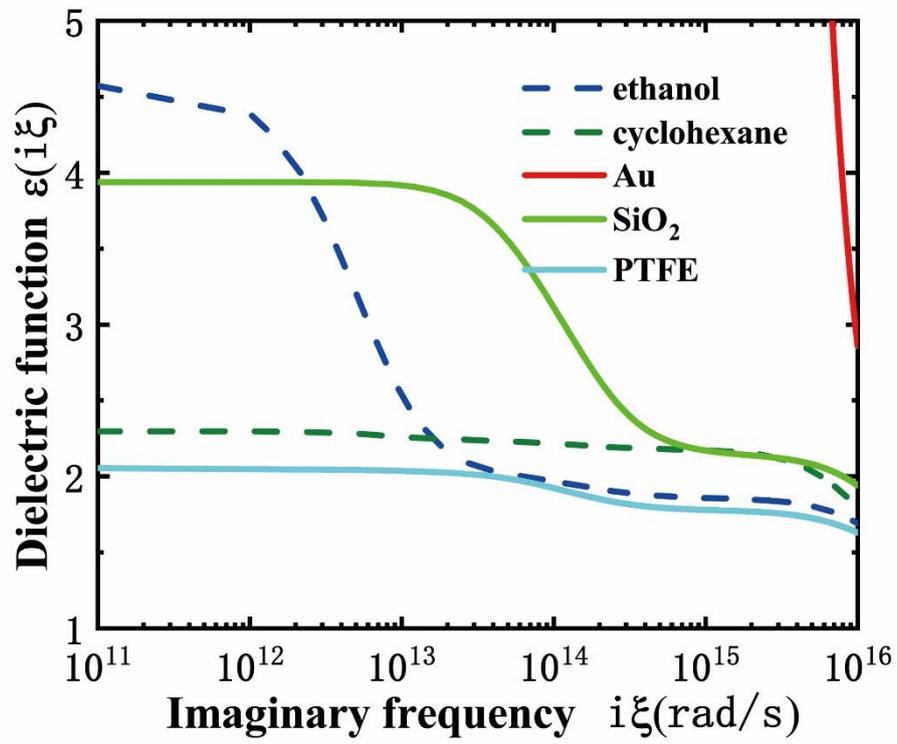

FIG. 5 (Color online) The dielectric functions of all experimental materials included versus the imaginary frequency